\documentclass[aps,prb,twocolumn]{revtex4}
\usepackage{graphicx}
\usepackage{bm}
\usepackage{amsmath}
\usepackage{bbold}
\usepackage[utf8]{inputenc}
\usepackage{color}
\usepackage{soul}
\usepackage[colorlinks=true, citecolor=blue]{hyperref}

\begin{document}
\title{A quantum teleportation experiment for undergraduate students}

\author{ S. Fedortchenko}
\email{serguei.fedortchenko@univ-paris-diderot.fr}

\affiliation{Laboratoire Mat\' eriaux et Ph\' enom\`enes Quantiques, Sorbonne Paris Cit\' e, Universit\' e Paris Diderot, CNRS UMR 7162, 75013, Paris, France}

\begin{abstract}
With the rapid progress of quantum information these recent years, it becomes more and more relevant to dedicate efforts in introducing this research topic to undergraduate students. However, as if in various fields of physics the theoretical learning is closely accompanied with experimental illustrations, such a teaching method can be expensive to organise for quantum information. We propose one way to circumvent these difficulties by using the recently made available 5-qubit quantum processor of IBM, in order to illustrate quantum information tasks, by actually implementing them on the superconducting chip through the web interface of the IBM Quantum Experience. We focus on the realization of a quantum teleportation protocol on this device, analyze the results and discuss the issues encountered, providing a complete analysis of this software as a pedagogical tool.
\end{abstract}
\vskip2pc

 
\vskip2pc 
\maketitle

\section{Introduction}

Quantum information studies information processing and transmission in systems governed by the principles of quantum physics. It has witnessed an impressive growth in the last thirty years, aiming to provide computational advantages over classical computers, or the investigation of new physics by simulating complex quantum systems.~\cite{QIreview} From the theoretical point of view, the first quantum information protocols emerged in the 1980s and the 1990s, such as quantum key distribution,~\cite{BB84} Deutsch-Jozsa's algorithm,~\cite{DeutschJozsa} quantum teleportation,~\cite{BennettTeleportation} or Shor's algorithm.~\cite{Shor} Meanwhile, on the experimental side, the progress occured much slowly, with nevertheless the generation of quantum correlations between two photons in 1981,~\cite{Aspect} and the demonstration of quantum cryptography in 1992.~\cite{BennettEXP} However, the experimental advances truly accelerated when a new method for the generation of entangled photon pairs became possible in 1995.~\cite{KwiatEXP} Indeed, shortly after, these tools were used to implement dense coding \cite{MattleEXP} and quantum teleportation,~\cite{BouwmeesterEXP} while atoms have also been proven useful to generate entangled states.~\cite{HagleyEXP} In later years, the technological progress enabled the experimental verification of various quantum information tasks.~\cite{SchmittEXP,ChuangEXP,GuldeEXP,RiebeEXP,VandersypenEXP,XuEXP} Much effort is put on their implementation in different physical systems, aiming the ultimate goal of building a quantum computer \cite{Ladd} or a quantum internet.~\cite{Pirandola} However, it is not yet clear which platform or platforms are the best candidates for these purposes.

Nevertheless, with the rapid advances of this branch of quantum physics, putting efforts in introducing it to undergraduate students is more pertinent than ever. First attempts to introduce the theory of quantum information and quantum computing can be traced back to the 1990s. \cite{Scarani,Candela} More recent efforts can be found, \cite{Gerjuoy,Benenti} as the use of original methods to explain Bell non-locality \cite{Jacobs} and quantum cryptography \cite{Svozil} for instance. Investigations on students' understanding of concepts related to quantum information and quantum computing have also been reported. \cite{Singh,Kohnle,DeVore} Proposals to illustrate quantum information through experiments exist, and are mainly based on photonic degrees of freedom, \cite{Funk,Brandt} in particular on the generation of entangled photon pairs. \cite{Holbrow,Dehlinger,Dehlinger2,Ourjoumtsev,Beck,Ashby} Some proposed experiments for undergraduate students use other kinds of physical systems, such as spins in nuclear magnetic resonance spectroscopy, \cite{Havel} or ultracold atoms confined in optical lattices. \cite{Foot} Yet, including experiments as illustrative examples in a teaching program can be difficult, and the reason for that is two-fold. First, quantum information experiments often require expensive equipment, and second, even the smallest-scale experiments are generally state-of-the-art research projects and require knowledge beyond the one achievable in a tutorial experiment in class. Even if some advanced undergraduate programs in prestigious institutes already include experiments such as the generation of entangled photon pairs,~\cite{IOGS} the theoretical side of quantum information is nonetheless the one mostly taught.

Surprisingly, this may change in the very near future. Indeed, recently IBM released a free web interface that allows one to run actual quantum information experiments on their 5-qubit quantum processor.~\cite{IBMQE} Called the IBM Quantum Experience (QE), this intuitive graphical interface permits the use of 5 transmon qubits \cite{TransmonPRA} to test quantum algorithms, which are sent to IBM to be implemented on their superconducting chip. This interface have already been used to conduct research experiments,~\cite{Takita,Alsina,Devitt,Berta,Rundle} and inspired a Python program that simulates the IBM QE interface and helps running actual experiments on it.~\cite{CorbettMoran} No a priori knowledge is needed in experimental physics for the anonymous user, that can easily access the results of his/her experiment. One of the attractions of the device is the rapidity with which algorithms are ran. Provided there is no queue (no other users sending requests), receiving the results of an experiment with 8192 realizations takes only a couple of minutes.

The IBM QE could then be used in classes to introduce to students the idea of quantum information with superconducting circuits, while making them run quantum algorithms and analyze the results, in order to illustrate the theory they study. Typically, the students could start by implementing basic operations such as simple one-qubit gates. In the QE, these gates are the identity gate $\hat{I}$, the $\hat{X}$, $\hat{Y}$, and $\hat{Z}$ Pauli gates, the Hadamard gate $\hat{H}$, the phase gate $\hat{S}$ and the $\hat{T}$ gate (for a definition of all the gates available in the IBM QE see the appendix \hyperref[appendixA]{A}). The last gate available is the two-qubit Controlled-NOT (CNOT) gate, allowing to create entanglement between two qubits. With all those tools, the students could afterwards test operations involving CNOT gates and create entanglement, from the simple algorithms to generate the well known two-qubit Bell states, or EPR states,~\cite{EPR} to the more advanced ones, generating the so-called Greenberger–Horne–Zeilinger (GHZ) states \cite{GHZ} for three, four or five qubits. Note that the IBM QE user guide already contain a certain number of protocols. The lesson could end with the implementation of a quantum teleportation protocol, with the associated theoretical calculations done in parallel to compare with the experimental results. Note that discrete variable quantum teleportation has been implemented with superconducting qubits in Refs.~[\onlinecite{Baur,Steffen}].

In this paper, we focus on the implementation of quantum teleportation with the interface provided by the IBM QE. We start by reviewing the theoretical calculations of the protocol in Section \ref{theory}. Then, in Section \ref{results} we show how to implement the protocol in the QE and show the results obtained. In Section \ref{issues} we detail the issues we faced when implementing the protocol and suggest how the QE could be improved to make the implementation better.

\section{The theory of quantum teleportation in discrete variables}
\label{theory}

Before showing how to actually implement a quantum teleportation with the IBM QE, we shall recall all the steps of the protocol along with the theoretical derivation that demonstrates the efficiency of it. The purpose of the protocol is the exchange of information between two parties, Alice and Bob, in the following manner: Alice has a qubit with an unknown state, and the goal of the protocol is to teleport this state on Bob's qubit, without obtaining any information about the state. In fact, it is impossible to obtain information about the state while teleporting it. The first step of a quantum teleportation is the generation of a Bell state, an entangled state shared by Alice and Bob,
\begin{equation}
\vert \psi_{\text{AB}} \rangle = \frac{\vert 0_{\text{A}} \rangle \otimes \vert 0_{\text{B}} \rangle + \vert 1_{\text{A}} \rangle \otimes \vert 1_{\text{B}} \rangle}{\sqrt{2}},
\label{bellstate}
\end{equation}
where the first qubit is owned by Alice and the second one by Bob. For the sake of simplicity, let us in the following omit the tensorial product, e.g., $\vert 0_{\text{A}} \rangle \otimes \vert 0_{\text{B}} \rangle \longrightarrow \vert 0_{\text{A}} 0_{\text{B}} \rangle$. Alice has an additional qubit $\vert \phi_i \rangle$, that she want to teleport to Bob. She must be able to do so irrespective of its state:
\begin{equation}
\vert \phi_{i} \rangle = \alpha \vert 0_{i} \rangle + \beta \vert 1_{i} \rangle,
\label{initialstate}
\end{equation}
where $\alpha$ and $\beta$ are the probability amplitudes of the $\vert 0_{i} \rangle$ and $\vert 1_{i} \rangle$ states. Hence, Alice has a pair of  qubits in her hands, the qubit A entangled with Bob's qubit and the qubit $i$, in an arbitrary state. She applies a CNOT gate to those two qubits, which unitary transformation can be written
\begin{equation}
    \hat{U}_{\text{CNOT}} = \frac{1}{\sqrt{2}} \left(
      \begin{array}{cccc}
        1 & 0 & 0 & 0 \\
        0 & 1 & 0 & 0 \\
        0 & 0 & 0 & 1 \\
        0 & 0 & 1 & 0
      \end{array} \right),
\end{equation}
in the basis $\{ \vert 0_{i}  0_{\text{A}} \rangle, \vert 0_{i} 1_{\text{A}} \rangle, \vert 1_{i} 0_{\text{A}} \rangle, \vert 1_{i} 1_{\text{A}} \rangle  \}$. The operation is controlled by the qubit $i$, that is, the state of qubit A is changed if the qubit $i$ is in state $\vert 1_{i} \rangle$, else A is not modified. After this gate, we have to describe the three qubits as a whole,
\begin{align}
\vert \Psi_{i\text{AB}} \rangle & = \hat{U}_{\text{CNOT}} \vert \phi_{i} \rangle \vert \psi_{\text{AB}} \rangle \nonumber \\
& = \frac{\alpha}{\sqrt{2}} \vert 0_{i} \rangle \Big( \vert 0_{\text{A}} 0_{\text{B}} \rangle + \vert 1_{\text{A}} 1_{\text{B}} \rangle \Big) + \nonumber \\
& + \frac{\beta}{\sqrt{2}} \vert 1_{i} \rangle \Big( \vert 1_{\text{A}} 0_{\text{B}} \rangle + \vert 0_{\text{A}} 1_{\text{B}} \rangle \Big).
\label{afterCNOT}
\end{align}
The next step is the implementation of a Hadamard gate on the qubit $i$,
\begin{equation}
    \hat{H} = \frac{1}{\sqrt{2}} \left(
      \begin{array}{cccc}
        1 & 1 \\
        1 & -1
      \end{array} \right),
\end{equation}
which gives the final state
\begin{align}
\hat{H} \vert \Psi_{i\text{AB}} \rangle & = \frac{1}{\sqrt{2}} \vert 0_{i} 0_{\text{A}} \rangle \big(  \alpha \vert 0_{\text{B}} \rangle + \beta \vert 1_{\text{B}} \rangle \big) + \nonumber \\
& + \frac{1}{\sqrt{2}} \vert 0_{i} 1_{\text{A}} \rangle \big(  \beta \vert 0_{\text{B}} \rangle + \alpha \vert 1_{\text{B}} \rangle \big) + \nonumber \\
& + \frac{1}{\sqrt{2}} \vert 1_{i} 0_{\text{A}} \rangle \big(  \alpha \vert 0_{\text{B}} \rangle - \beta \vert 1_{\text{B}} \rangle \big) + \nonumber \\
& + \frac{1}{\sqrt{2}} \vert 1_{i} 1_{\text{A}} \rangle \big(  - \beta \vert 0_{\text{B}} \rangle + \alpha \vert 1_{\text{B}} \rangle \big).
\label{afterH}
\end{align}
The following step of the protocol is the measurement of qubits $i$ and A by Alice in the $Z$ basis, which projects Bob's qubit into one of the four possible states in Eq. (\ref{afterH}). The results of both measurements are sent to Bob through a classical channel, and according to those results, Bob has to apply to his qubit one of the four following gates: $\hat{I}$, $\hat{X}$, $\hat{Z}$, or $\hat{Y}$. This last operation completes the protocol.

\section{Experimental implementation of the protocol and results}
\label{results}

\begin{figure}[t!]
\centering
\includegraphics[width=0.50\textwidth]{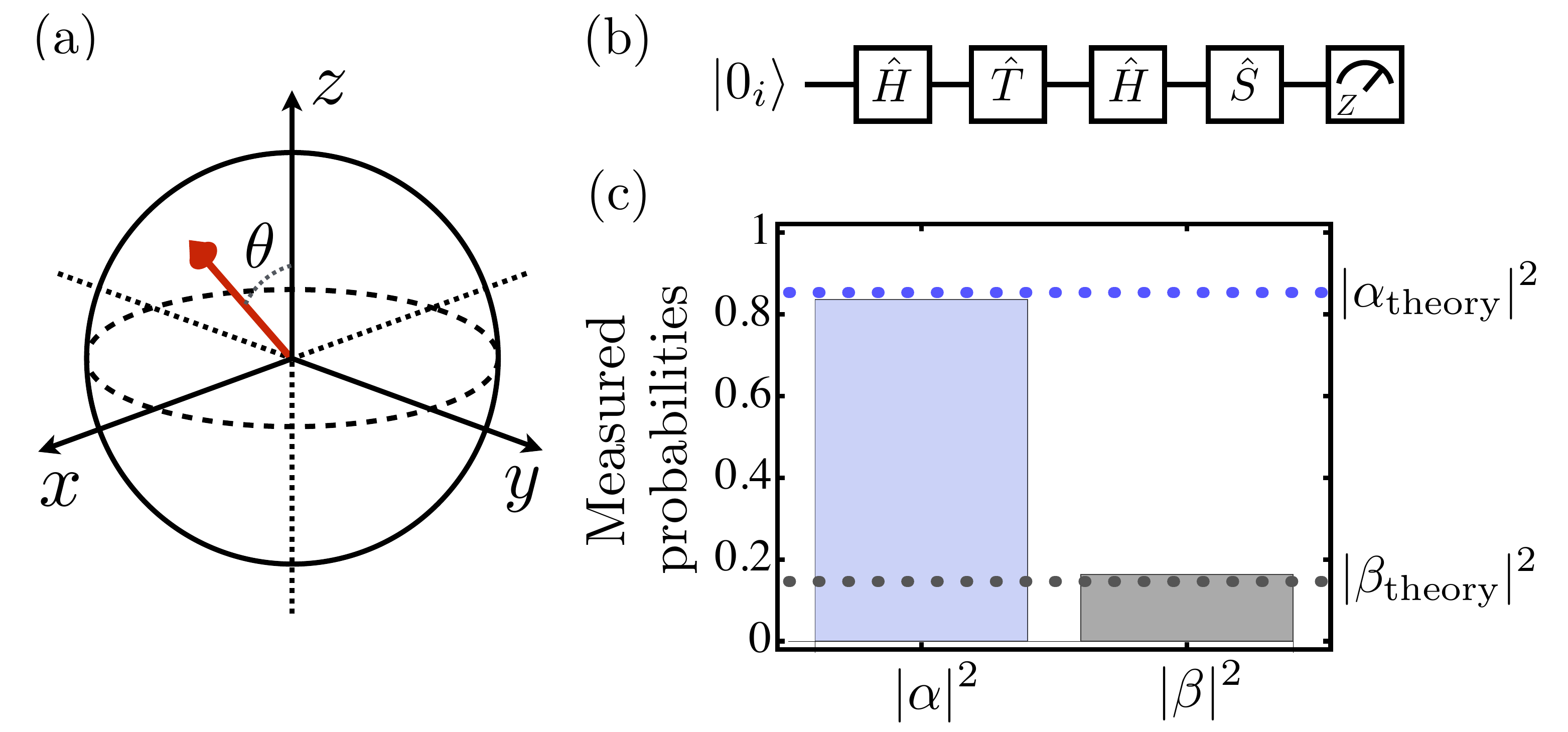}
\caption{(a) Ideal representation of the teleported state in the Bloch sphere. (b) Circuit used in the IBM QE to generate this initial state and measure its probabilities. The qubit used in the interface is qubit 1. (c) The bars indicate the measured probabilities (in the $Z$ basis) of the generated initial state following the protocol given in (b). The dotted lines are the theoretical values for these probabilities. The results are averaged over 8192 realizations.}
\label{FigureInitial}
\end{figure}

We shall now move to the experimental implementation of the protocol using the IBM QE interface. The architecture of their quantum processor is based on transmon qubits.~\cite{TransmonPRA} A transmon qubit is a charge qubit, meaning that the quantized levels used to define the qubit are charge levels. These are made of electrical circuits based on superconducting materials, in which are inserted weak links, known as Josephson junctions. Weak links are layers of non-superconducting materials such as insulators.

We chose to teleport a state where $\alpha=\cos{(\theta/2)}$ and $\beta=\sin{(\theta/2)}$, with $\theta=\pi/4$. A representation of this state in the Bloch sphere is given in Fig.~\ref{FigureInitial}(a). Four single-qubit gates are needed to generate this state, and are specified in Fig.~\ref{FigureInitial}(b), which shows the associated circuit. For a theoretical derivation of this state using the gates of the circuit, see the appendix \hyperref[appendixB]{B}. We implemented this circuit in the IBM QE, and the measured probabilities for this initial state are shown in Fig.~\ref{FigureInitial}(c). One can notice good agreement between theory and experiment. 

\begin{figure}[t!]
\centering
\includegraphics[width=0.47\textwidth]{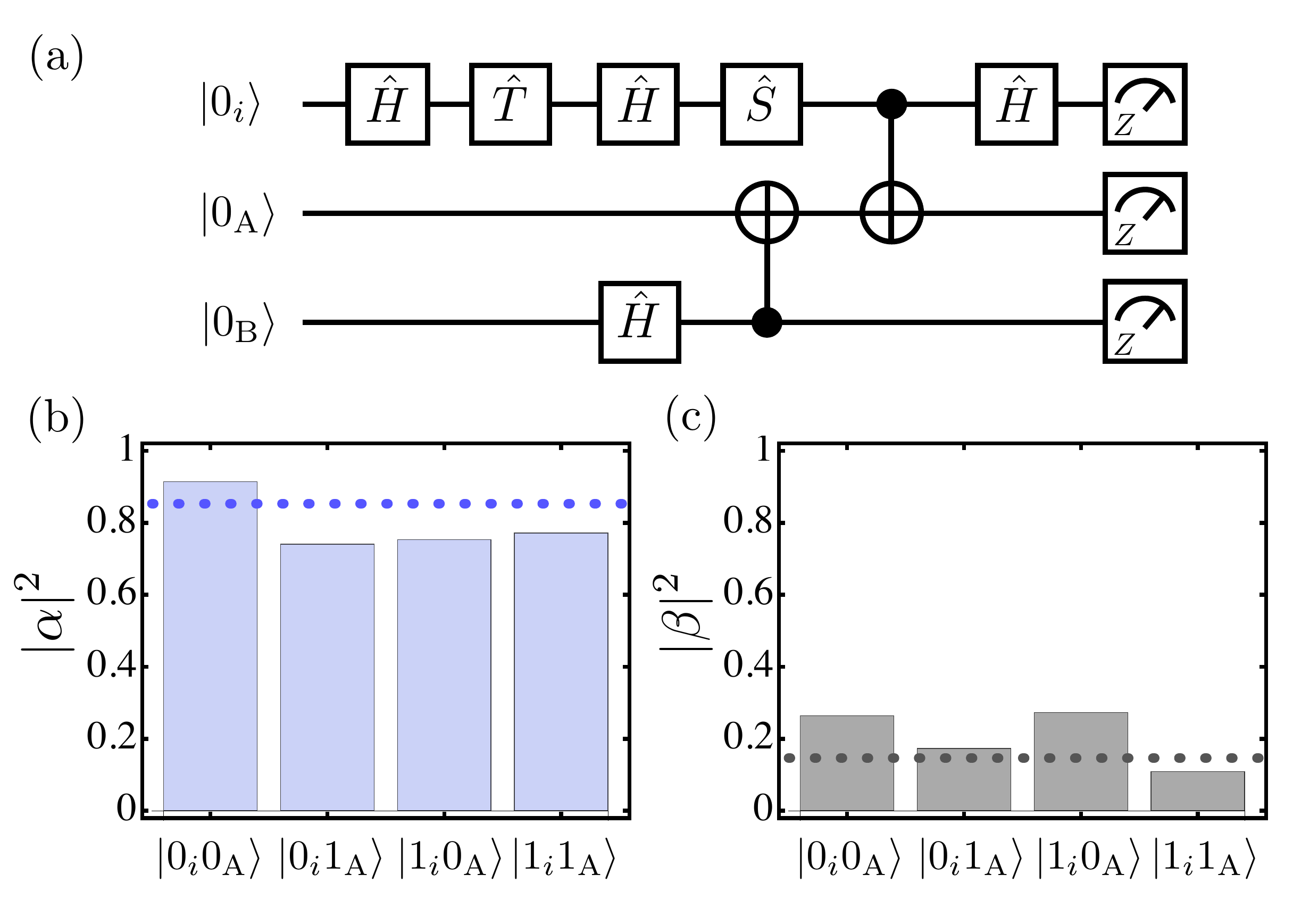}
\caption{(a) Circuit used in the IBM QE to implement the quantum teleportation. The qubits used in the interface are qubits 1, 2 and 3. The gates joining two qubits are CNOT gates. (b)-(c) Experimental results, populations measured in the $Z$ basis. (b) The bars indicate the measured probability $\vert \alpha \vert^2$ of Bob's final state, as a function of Alice's measurement outcomes. The dotted line is the theoretical value $\vert \alpha_{\text{theory}} \vert^2$. (c) The bars indicate the measured probability $\vert \beta \vert^2$ of Bob's final state, as a function of Alice's measurement outcomes. The dotted line is the theoretical value $\vert \beta_{\text{theory}} \vert^2$. The results are averaged over 8192 realizations.}
\label{FigureResults}
\end{figure}

After the characterisation of the initial state, we implemented the quantum teleportation using the circuit shown in Fig.~\ref{FigureResults}(a). In Fig.~\ref{FigureResults}(b) and \ref{FigureResults}(c), we respectively show the measured populations $\vert \alpha \vert^2$ and $\vert \beta \vert^2$ of Bob's final state, as functions of Alice's measurements outcomes.~\cite{Data} While the results are more noisy than in Fig.~\ref{FigureInitial}(c), they still reasonably follow the theoretical predictions. 

As an exercise for students, the quantum teleportation circuit has the advantage of being relatively simple, and yet permits to realise one of the most famous protocols in quantum information and address the potential problematic issued when running an experiment either in the lab or in such an interface.

\section{Discussion on the issues encountered}
\label{issues}

We will now detail the issues faced when we implemented the protocol, inherent of the IBM QE, and possibly inconvenient for many quantum algorithms that could be ran using this interface. Note that it may be a good exercise for students to try to understand and to guess by themselves what are the issues and limitations of the IBM QE. The first limitation is the impossibility of intuitively accessing the coherences of entangled states. Indeed, for an entangled state, the only observables that can be measured are the populations through a measurement in the $Z$ basis. This means that if one is given the results in Fig.~\ref{FigureResults}(b) and \ref{FigureResults}(c) without knowing what circuit was used, there would be no way to know if the populations are for a state with quantum coherence or a mixed state. In other words, the full tomography of an entangled state is not available as an accessible tool. Note that in a very recent work, it is shown how to reconstruct the spin Wigner function of Bell states.~\cite{Rundle}

Nevertheless, coherence can be easily measured for a single-qubit state that is separable from any other qubit. This is achieved by performing what is called in the IBM QE a Bloch sphere measurement. This measurement gives the vector of the state in the Bloch sphere. We could have shown this measured Bloch vector in Fig.~\ref{FigureInitial}(c) for the initial state of Alice, but we have shown the populations of the state instead, to be consistent with the way the results are shown in Fig.~\ref{FigureResults}(b) and \ref{FigureResults}(c). We cannot use Bloch sphere measurements in the teleportation circuit in Fig.~\ref{FigureResults}(a), because otherwise we would obtain vectors of mixed states, due to the entanglement they shared before the measurement.

This last comment points out another limitation present in the IBM QE device, specifically problematic for the protocol we implemented. In fact, the circuit shown in Fig.~\ref{FigureResults}(a) may be rightfully seen as incomplete for quantum teleportation. Indeed, the three qubits are measured at the same time, while Alice should first measure her two qubits, then send the results to Bob through a classical channel, who performs on his qubit a last operation conditioned by the results he received, as explained in Section \ref{theory}. Conditioning the application of gates by the outcomes of measurements is not possible in the IBM QE, and one can only obtain the populations of all the possible states in the $Z$ basis. In our experiment, we applied this last step as a post-selection in the results, meaning that the application of the last gate on Bob's qubit is assumed to be performed with perfect fidelity.

Another issue to be addressed is on the spacial separation between Alice and Bob. In fact, a crucial detail in a quantum teleportation protocol is that both parties should be spatially distant, since it is a communication protocol. This criterion cannot be achieved in the IBM QE, neither was it the case in the previous works studying quantum teleportation between superconducting qubits.~\cite{Baur,Steffen} Note that this problem can be overcome in other physical systems, as in Refs.~[\onlinecite{Aspect,BouwmeesterEXP}], where the protocol is implemented between two travelling photons.

Additionally, because of the limitations detailed before, we could not use the standard benchmark for a quantum teleportation protocol, called the fidelity. It is defined as $\mathcal{F}=\langle   \phi_{i}^{\text{theory}} \vert  \rho_{\text{B}}^{\text{exp}}  \vert \phi_{i}^{\text{theory}} \rangle$, where $\vert \phi_{i}^{\text{theory}} \rangle$ is the teleported state in the perfect theoretical case, and where $\rho_{\text{B}}^{\text{exp}}$ is the density matrix of the final state of Bob's qubit in the experiment. In order to apply this criterion, all the limitations, except the distance between the qubits, must be improved in an updated version of the IBM QE.

\section{Conclusion}
\label{conclusion}

We have shown how undergraduate students can easily implement quantum information protocols, and in particular a quantum teleportation protocol, using the IBM QE interface. Reserved only to a small number of experts not so long ago,~\cite{Baur,Steffen} quantum teleportation with superconducting qubits can now be tested by any person interested in quantum information, and more generally, in quantum mechanics. Though its implementation is not perfect due to limitations of the device, this interface is a nice playground that can help illustrating key concepts in quantum information. Since making undergraduate students run such experiments is usually difficult and expensive for introductory courses in this field, we believe the IBM QE could become a useful tool for teachers willing to immerse their students in it. Hopefully, an improvement of the device could lead to more extensive use of this interface, not only to teach quantum information, but to conduct more research as well, as started in Refs.~[\onlinecite{Takita,Alsina,Devitt,Berta,Rundle}].

\section*{Acknowledgments}
\label{acknowledgments}

This work was supported by the ANR COMB project, grant ANR-13-BS04-0014 of the French Agence Nationale de la Recherche. The author thank IBM for the opportunity of using a 5-qubit quantum processor through the IBM QE interface. The discussions and opinions developed in this paper are only those of the author and do not reflect the opinions of IBM employees. The author also thanks P. Milman, A. Keller, and T. Coudreau whose comments helped to improve the manuscript.

\section*{Appendix A: Definition of the quantum gates available in the IBM QE}
\label{appendixA}

Here we define all the gates available in IBM QE interface, by giving their expression in a matrix form. For gates applied to one qubit, their expression is given in the basis $\{ \vert 0 \rangle, \vert 1 \rangle  \}$, while the two qubit gate is written in the following basis $\{ \vert 0  0 \rangle, \vert 0 1 \rangle, \vert 1 0\rangle, \vert 1 1 \rangle  \}$. The identity gate and the Pauli gates are defined as,
\begin{align}
    \hat{I} & = \left(
      \begin{array}{cccc}
        1 & 0 \\
        0 & 1
      \end{array} \right), \quad \hat{X} = \left(
      \begin{array}{cccc}
        0 & 1 \\
        1 & 0
      \end{array} \right), \nonumber \\
\hat{Y} & = \left(
      \begin{array}{cccc}
        0 & -i \\
        i & 0
      \end{array} \right), \quad \hat{Z} = \left(
      \begin{array}{cccc}
        1 & 0 \\
        0 & -1
      \end{array} \right).  
\end{align}
The phase gate and the $\hat{T}$ gate are defined as,
\begin{align}    
      \hat{S} & =  \left(
      \begin{array}{cccc}
        1 & 0 \\
        0 & i
      \end{array} \right), \quad \hat{T} = \left(
      \begin{array}{cccc}
        1 & 0 \\
        0 & e^{i\pi/4}
      \end{array} \right).
      \label{SandT}  
\end{align}
The Hadamard gate, allowing to create a coherent superposition, is written as
\begin{equation}
      \hat{H} = \frac{1}{\sqrt{2}} \left(
      \begin{array}{cccc}
        1 & 1 \\
        1 & -1
      \end{array} \right).  
\end{equation}
Finally, the Controlled-NOT (CNOT) gate, that creates entanglement between two qubits, can be written as
\begin{equation}
    \hat{U}_{\text{CNOT}} = \frac{1}{\sqrt{2}} \left(
      \begin{array}{cccc}
        1 & 0 & 0 & 0 \\
        0 & 1 & 0 & 0 \\
        0 & 0 & 0 & 1 \\
        0 & 0 & 1 & 0
      \end{array} \right).
\end{equation}

\section*{Appendix B: Derivation of Alice's initial state}
\label{appendixB}

Here we show how to derive the probability amplitudes of Alice's initial state $\vert \phi_{i} \rangle$, using the circuit in Fig.~\ref{FigureInitial}(b). The qubit starts in $\vert 0_{i} \rangle$, and we first apply a Hadamard gate to it, which gives us,
\begin{equation}
\hat{H} \vert 0_{i} \rangle = \frac{ \vert 0_{i} \rangle +  \vert 1_{i} \rangle }{\sqrt{2}}.
\label{1}
\end{equation}
Then, we use a $\hat{T}$ gate, which change the state of the qubit into
\begin{equation}
\hat{T} \hat{H} \vert 0_{i} \rangle = \frac{  \vert 0_{i} \rangle +  e^{i\pi/4} \vert 1_{i} \rangle }{\sqrt{2}}.
\label{3}
\end{equation}
This is followed by another Hadamard gate,
\begin{equation}
\hat{H} \hat{T} \hat{H} \vert 0_{i} \rangle = \frac{ (1+e^{i\pi/4}) \vert 0_{i} \rangle +  (1-e^{i\pi/4}) \vert 1_{i} \rangle }{2},
\label{4}
\end{equation}
and finally an $\hat{S}$ gate
\begin{equation}
    \hat{S} = \left(
      \begin{array}{cccc}
        1 & 0 \\
        0 & i
      \end{array} \right),
\end{equation}
giving
\begin{equation}
\vert \phi_{i} \rangle = \frac{ (1+e^{i\pi/4}) \vert 0_{i} \rangle +  i(1-e^{i\pi/4}) \vert 1_{i} \rangle }{2},
\label{5}
\end{equation}
which can be simplified into
\begin{equation}
\vert \phi_{i} \rangle = e^{i\pi/8} \Big( \cos{(\pi/8)} \vert 0_{i} \rangle +  \sin{(\pi/8)} \vert 1_{i} \rangle \Big),
\label{6}
\end{equation}
where $e^{i\pi/8}$ is simply a global phase.


\begin{thebibliography}{99}

\bibitem{QIreview} C. H. Bennett and D. P. DiVincenzo, Nature {\bf 404}, 247–255 (2000).


\bibitem{BB84} C. H. Bennett and G. Brassard, in Proceedings of the IEEE International Conference on Computers, Systems and Signal Processing, Bangalore, India, 1984 (IEEE, New York, 1984), pp. 175–179; IBM Tech. Discl. Bull. {\bf 28}, 3153–3163 (1985).


\bibitem{DeutschJozsa} D. Deutsch and R. Jozsa, Proc. R. Soc. London, Ser. A {\bf 439}, 553 (1992).


\bibitem{BennettTeleportation} C. H. Bennett, G. Brassard, C. Crepeau, R. Jozsa, A. Peres, and W. K. Wootters, Phys. Rev. Lett. {\bf 70}, 1895 (1993).

\bibitem{Shor}  P. Shor, Proc. 35th Ann. Symp. Found. Comp. Sci. (IEEE Comp. Soc. Press, Los Alamitos, California, 1994), p. 124.


\bibitem{Aspect} A. Aspect, P. Grangier, and G. Roger, Phys. Rev. Lett. {\bf 47}, 460 (1981).


\bibitem{BennettEXP} C. H. Bennett, F. Bessette, G. Brassard, L. Salvail, and J. Smolin, J. Cryptol. {\bf 5}, 3 (1992).

\bibitem{KwiatEXP} P. G. Kwiat, K. Mattle, H. Weinfurter, A. Zeilinger, A. V. Sergienko, and Y. Shih, Phys. Rev. Lett. {\bf 75}, 4337 (1995).

\bibitem{MattleEXP} K. Mattle, H. Weinfurter, P. G. Kwiat, and A. Zeilinger, Phys. Rev. Lett. {\bf 76}, 4656 (1996).


\bibitem{BouwmeesterEXP} D. Bouwmeester, J.-W. Pan, K. Mattle, M. Eibl, H. Weinfurter, and A. Zeilinger, Nature {\bf 390}, 575-579 (1997).


\bibitem{HagleyEXP} E. Hagley, X. Maître, G. Nogues, C. Wunderlich, M. Brune, J. M. Raimond, and S. Haroche, Phys. Rev. Lett. {\bf 79}, 1 (1997).


\bibitem{ChuangEXP} I. I. Chuang, I. M. K. Vandersypen, X. Zhou, D. W. Leung, and S. Lloyd, Nature {\bf 393}, 143-146 (1998).


\bibitem{VandersypenEXP} L. M. K. Vandersypen, M. Steffen, G. Breyta, C. S. Yannoni, M. H. Sherwood, and I. L. Chuang, Nature {\bf 414}, 883-887 (2001).


\bibitem{GuldeEXP} S. Gulde, M. Riebe, G. P. T. Lancaster, C. Becher, J. Eschner, H. H\"affner, F. Schmidt-Kaler, Isaac L. Chuang, and R. Blatt, Nature {\bf 421}, 48-50 (2003).


\bibitem{RiebeEXP} M. Riebe, H. H\"affner, C. F. Roos, W. H\"ansel, J. Benhelm, G. P. T. Lancaster, T. W. K\"orber, C. Becher, F. Schmidt-Kaler, D. F. V. James, and R. Blatt, Nature {\bf 429}, 734-737 (2004).


\bibitem{SchmittEXP} T. Schmitt-Manderbach, H. Weier, M. F\"urst, R. Ursin, F. Tiefenbacher, T. Scheidl, J. Perdigues, Z. Sodnik, C. Kurtsiefer, J. G. Rarity, A. Zeilinger, and H. Weinfurter, Phys. Rev. Lett. {\bf 98}, 010504 (2007).


\bibitem{XuEXP} N. Xu, J. Zhu, D. Lu, X. Zhou, X. Peng, and J. Du, Phys. Rev. Lett. {\bf 108}, 130501 (2012).


\bibitem{Ladd} T. D. Ladd, F. Jelezko, R. Laflamme, Y. Nakamura, C. Monroen and J. L. O’Brien, Nature {\bf 464}, 45-53 (2010).

\bibitem{Pirandola} S. Pirandola and S. L. Braunstein, Nature {\bf 532}, 169 (2016).


\bibitem{Scarani} V. Scarani, Am. J. Phys. {\bf 66}, 956 (1998). 

\bibitem{Candela} D. Candela, Am. J. Phys. {\bf 67}, 434 (1999). 


\bibitem{Gerjuoy} E. Gerjuoy, Am. J. Phys. {\bf 73}, 521 (2005). 


\bibitem{Benenti} G. Benenti, and G. Strini, Am. J. Phys. {\bf 76}, 657 (2008). 

\bibitem{Jacobs} K. Jacobs and H. M. Wiseman, Am. J. Phys. {\bf 73}, 932 (2005). 

\bibitem{Svozil} K. Svozil, Am. J. Phys. {\bf 74}, 800 (2006). 

\bibitem{Singh} C. Singh, AIP Conf. Proc. {\bf 883}, 42 (2007).

\bibitem{Kohnle} A. Kohnle and E. Deffebach, Phys. Educ. Res. Conf. Proc. (edited by A.D. Churukian, D. Jones, and L. Ding), p. 171-174 (2015). 


\bibitem{DeVore} S. DeVore and C. Singh, arXiv:1601.01301 (2016). 


\bibitem{Funk} A. C. Funk and M. Beck, Am. J. Phys. {\bf 65}, 492 (1997).

\bibitem{Brandt} H. E. Brandt, Am. J. Phys. {\bf 67}, 434 (1999). 

\bibitem{Holbrow} C. H. Holbrow, E. Galvez, and M. E. Parks, Am. J. Phys. {\bf 70}, 260 (2002). 

\bibitem{Dehlinger} D. Dehlinger and M. W. Mitchell, Am. J. Phys. {\bf 70}, 898 (2002). 

\bibitem{Dehlinger2} D. Dehlinger and M. W. Mitchell, Am. J. Phys. {\bf 70}, 903 (2002).


\bibitem{Ourjoumtsev} A. Ourjoumtsev, M.-C. Dheur, T. Avignon, and L. Jacubowiez, Eur. J. Phys. {\bf 36}, 065034 (2015). 


\bibitem{Beck} M. N. Beck, and M. Beck, Am. J. Phys. {\bf 84}, 87 (2016). 

\bibitem{Ashby} J. M. Ashby, P. D. Schwarz, and M. Schlosshauer, Am. J. Phys. {\bf 84}, 95 (2016). 


\bibitem{Havel} T. F. Havel, D. G. Cory, S. Lloyd, N. Boulant, E. M. Fortunato, M. A. Pravia, G. Teklemariam, Y. S. Weinstein, A. Bhattacharyya and J. Hou, Am. J. Phys. {\bf 70}, 345 (2002). 


\bibitem{Foot} C. J. Foot and M. D. Shotter, Am. J. Phys. {\bf 79}, 762 (2011).


\bibitem{IOGS} Institut d'Optique Graduate School, Universit\'e Paris Sud, Palaiseau, France. URL \url{https://www.institutoptique.fr/en/Formation/Ingenieur-SupOptique2/Travaux-Pratiques/Physique-quantique-atomique-nanophysique}. 


\bibitem{IBMQE} IBM. The Quantum Experience. URL \url{http://www.research.ibm.com/quantum/}.


\bibitem{TransmonPRA} J. Koch, T. M. Yu, J. Gambetta, A. A. Houck, D. I. Schuster, J. Majer, A. Blais, M. H. Devoret, S. M. Girvin, and R. J. Schoelkopf, Phys. Rev. A {\bf 76}, 042319 (2007).



\bibitem{Takita} M. Takita, A. D. C\'orcoles, E. Magesan, B. Abdo,
M. Brink, A. Cross, J. M. Chow, and J. M. Gambetta, arXiv:1605.01351v2 (2016). 


\bibitem{Alsina} D. Alsina, J. I. Latorre, arXiv:1605.04220v1 (2016). 


\bibitem{Devitt} S. J. Devitt, arXiv:1605.05709v3 (2016). 


\bibitem{Berta} M. Berta, S. Wehner, and M. M. Wilde, arXiv:1511.00267v2 (2016). 


\bibitem{Rundle} R. P. Rundle, T. Tilma, J. H. Samson, and M. J. Everitt, arXiv:1605.08922v2 (2016). 


\bibitem{CorbettMoran} C. Corbett Moran, arXiv:1605.05709v3 (2016). 


\bibitem{EPR} A. Einstein, B. Podolsky, and N. Rosen, Phys. Rev. {\bf 47}, 777 (1935).


\bibitem{GHZ} D. M. Greenberger, M. Horne, A. Zeilinger, "Going beyond Bell's theorem" in \textit{Bell's Theorem, Quantum Theory, and Conceptions of the Universe}, edited by M. Kafatos (Kluwer Academic, Dordrecht, The Netherlands, 1989) pp. 73-76.



\bibitem{Baur} M. Baur, A. Fedorov, L. Steffen, S. Filipp, M. P. da Silva, and A. Wallraff, Phys. Rev. Lett. {\bf 108}, 040502 (2012).

\bibitem{Steffen} L. Steffen, Y. Salathe, M. Oppliger, P. Kurpiers,	 M. Baur, C. Lang, C. Eichler, G. Puebla-Hellmann, A. Fedorov, and A. Wallraff, Nature {\bf 500}, 319–322 (2013).









\bibitem{Data} Experimental data available upon request.


\end{thebibliography}
\end{document}